\documentclass[%
 reprint,
superscriptaddress,
nofootinbib,
 amsmath,amssymb,
 aps,
]{revtex4-1}
\usepackage{graphicx}
\usepackage{dcolumn}
\usepackage{bm}
\usepackage{hyperref}

\usepackage{xcolor}
\usepackage{array}
\usepackage{caption} 

\begin{document}

\preprint{APS/123-QED}

\title{Periodic Orbits and Gravitational Wave Radiation of Black Hole in EGB gravity
}%

\author{Liping Meng}

\author{Zhaoyi Xu}


 \author{Meirong Tang}
 \email{Electronic address: tangmr@gzu.edu.cn(Corresponding author)
}
\affiliation{%
College of Physics, Guizhou University,\\
 Guiyang 550025, China
}%





\date{\today}

\begin{abstract}
This paper investigates the orbital dynamics and gravitational wave radiation characteristics of neutral test particles around a static spherically symmetric charged black hole (BH) in 4D Einstein-Gauss-Bonnet (4D-EGB) gravity theory.
 We analyze the dependence of the marginally bound orbit (MBO) and the innermost stable circular orbit (ISCO) on the Gauss-Bonnet coupling parameter $\alpha$ and charge $Q$. The results indicate that the orbital radius, angular momentum, and energy all decrease with increasing $\alpha$ or $Q$, with the corresponding bound orbit region shifting leftward in the $(E, L)$ parameter space. By combining observational data from the BH shadows of M87* and Sgr A* as well as the orbital precession of the S2 star, we constrain the model parameters and find that existing observations can limit the ranges of $\alpha$ and $Q$ to a certain extent. Furthermore, we investigate the characteristics of periodic orbits corresponding to different rational numbers $q$ and the gravitational waveforms they excite, finding that variations in $\alpha$ and $Q$ can lead to distinguishable differences in periodic orbit structures and gravitational wave phases. This study contributes to understanding the effects of Gauss-Bonnet corrections on BH spacetimes, and the results may provide theoretical references for future gravitational wave observations of extreme mass ratio inspirals (EMRIs).

\begin{description}
\item[Keywords]
4D-EGB Gravity Theory, Gravitational Waves, Periodic Orbits, Charged Black Hole
\end{description}
\end{abstract}


\maketitle

\section{Introduction}\label{1.0}
Since the advent of general relativity(GR), research on this theory of gravity has been endless. Currently, GR has been confirmed by multiple observations and experiments, for instance gravitational waves and BH shadows, among others\cite{LIGOScientific:2016aoc,EventHorizonTelescope:2019dse,Davis:1982gc,Tonry:1998ci,Campbell:1983ck}. However, under extreme conditions, issues such as spacetime singularities\cite{Hawking:1970zqf,Penrose:1964wq} and the BH information paradox\cite{Hawking:2015qqa} indicate the limitations of GR. These theoretical and observational problems have promoted researchers to explore alternative theories. So far, the main alternative theories to GR include theoretical models such as string theory\cite{Polchinski:1998rq,Polchinski:1998rr}, loop quantum gravity\cite{Thiemann:2002nj,Han:2005km,Ashtekar:2006wn}, scalar-tensor theories of gravity\cite{Lecoeur:2024kwe,Bekenstein:1974sf,Callan:1970ze}, and $f(R)$ gravity\cite{Nojiri:2006ri,Hu:2007nk}. Among these numerous alternative theoretical models, the Einstein-Gauss-Bonnet (EGB) theory\cite{Lovelock:1971yv,Lovelock:1972vz,Lanczos:1938sf,Fernandes:2022zrq} holds a significant position in modern gravitational theory research due to its unique theoretical construction and mathematical properties. Specifically, only second derivatives of the metric appear in this theory's field equations (the same as GR), and beyond GR, no new fundamental fields are needed.

However, in four-dimensional spacetime, the Gauss-Bonnet term is a topological invariant, and according to the Lovelock theorem, its variation does not contribute dynamically to the field equations. To circumvent this limitation, Glavan and Lin \cite{Glavan:2019inb}  proposed a singular rescaling of the coupling constant $\alpha \to \alpha/(D-4)$ and taking the $D \to 4$ limit at the level of the action, thereby enabling the Gauss-Bonnet term to produce non-trivial effects and establishing the 4D-EGB gravity theory. This approach has sparked extensive debate due to the mathematical ambiguity in the definition of the limiting procedure\cite{Gurses:2020ofy,Mahapatra:2020rds,Shu:2020cjw}. 
To address this issue, several studies have established mathematically well-defined four-dimensional effective field theory frameworks through dimensional reduction, counter-term regularization, and embedding into Horndeski scalar-tensor theory \cite{Aoki:2020lig,Aoki:2020iwm,Fernandes:2020nbq,Hennigar:2020lsl}. 
It is worth emphasizing that the static spherically symmetric solutions obtained from these regularized theories are completely consistent in metric form with those derived from the original $D \to 4$ limiting method\cite{Lu:2020iav,Mann:1992ar,Kobayashi:2020wqy}. 
This implies that the spherically symmetric BH solutions obtained via the $D \to 4$ limit are also exact solutions of these well-defined theories, thereby providing a solid theoretical foundation for their physical applications in four-dimensional spacetime. Based on this, 4D-EGB gravity has become a research hotspot in the field of modified gravity, with related works extensively covering BH physics, cosmology, and weak-field tests \cite{Hassannejad:2023lrp,Gammon:2022bfu,Kumar:2023ijg,Xie:2023tjc,Panyasiripan:2024iww,Xie:2023tjc}. In the study of BH solutions, Fernandes \cite{Fernandes:2020rpa} derived the static spherically symmetric charged BH solution within the 4D-EGB theoretical framework. In this paper, we take this spacetime as the background to systematically investigate the orbital dynamics and gravitational wave radiation properties of charged BH in 4D-EGB gravity theory, aiming to reveal the effects of Gauss-Bonnet corrections on BH spacetime structure and its observational signatures.

In BH spacetimes, stellar-mass particle orbits are studied via timelike geodesics\cite{Kostic:2012zw,Hackmann:2008tu,Fujita:2009bp}. Furthermore, current research indicates that the bound orbits of particles exhibit distinct characteristics\cite{Glampedakis:2002ya,Barack:2003fp,GRAVITY:2020gka}. Therefore, studying the bound orbits of particles is crucial for understanding the physical properties and spacetime structural characteristics of BHs. Moreover, as an important type of bound orbit, periodic orbits are crucial for exploring gravitational wave radiation and BH orbital dynamics\cite{Lake:2003tr,Glampedakis:2002ya}. Particularly after the successful detection of GWs\cite{LIGOScientific:2016aoc}, the research of periodic orbits has attracted great interest. In light of this, Levin et al. put forward a classification of periodic orbits\cite{Levin:2008mq}. This classification method introduces three integers, $z$, $w$, and $v$, which represent the zoom number, whirl number, and vertex number of the particle orbit, respectively. Based on this, a rational number $q = w + v/z$ is defined as the classification feature, and a one-to-one correspondence between the rational number $q$ and different periodic orbits is established, thereby achieving the purpose of classification. Currently, this classification method has gained significant recognition in relevant research fields, as seen in literature such as\cite{Zhao:2024exh,Wang:2025hla,Huang:2024oli,Meng:2024cnq}.

Furthermore, in binary systems, particularly in EMRIs composed of a supermassive BH and a small compact object, the smaller body spirals slowly toward the central BH along geodesics due to energy and angular momentum dissipation caused by gravitational wave radiation. Under the adiabatic approximation, the inspiral process can be described as a gradual transition of the particle through a series of quasi-periodic orbits, with each stage of the periodic orbit regarded as an instantaneous quasi-steady state. Differences in spacetime metrics under different gravitational theory frameworks systematically alter the dynamical properties of these periodic orbits, thereby leaving observable imprints in the evolution of gravitational wave signals, which provides a feasible approach for probing gravitational theories through gravitational wave detection. With the advancement of space-based gravitational wave detector missions such as LISA, EMRI systems have been recognized as one of the key sources for probing the spacetime properties in strong gravitational field regions \cite{Berry:2019wgg,Babak:2017tow}. In this paper, within the framework of 4D-EGB gravity theory, we systematically investigate the periodic orbits and gravitational wave radiation characteristics of neutral test particles around a charged BH. By analyzing the effects of the Gauss-Bonnet coupling parameter $\alpha$ and charge $Q$ on the geometric morphology of periodic orbits and gravitational waveforms, we aim to provide some theoretical references for distinguishing different types of BHs in future space-based gravitational wave detection.

The paper is structured as follows: In Section \ref{2.0} introduces the basic properties of the charged BH under 4D-EGB gravity theory, and based on geodesics, analyzes the characteristics of the particle's MBO and ISCO. In Section \ref{3.0}, we impose strict physical constraints on the parameter $\alpha$ and charge $Q$ by combining BH shadow effects and precession phenomena. In Sections \ref{4.0} and \ref{5.0}, we study the periodic trajectories of the particle around the charged BH and the corresponding gravitational wave radiation, respectively. Section \ref{6.0} provides a summary. Finally, it is important to emphasize that this study adopts natural units where $c=G=1$.

\section{Background of the BH and geodesics, MBO, and ISCO in 4D-EGB gravity}\label{2.0}
\subsection{Background of the BH}\label{2.1}

The static spherically symmetric charged BH solution given by Fernandes can be described by the following action
\begin{equation}
	S=\frac{1}{16\pi}\int{d^Dx\sqrt{-g}}\left[R+\frac{\alpha}{D-4}\mathcal{G}-F_{\mu\nu}F^{\mu\nu}\right],\label{eq1}
\end{equation}
where the Gauss-Bonnet term is
\begin{equation}
	\mathcal{G}=R^2-4R_{\mu\nu}R^{\mu\nu}+R_{\mu\nu\rho\sigma}R^{\mu\nu\rho\sigma}.\label{eq2}
\end{equation}
Here $D$ is the spacetime dimension, $\alpha$ is the Gauss-Bonnet coupling parameter, and $F_{\mu\nu}=\partial_\mu A_\nu-\partial_\nu A_\mu$ is the Maxwell tensor.

Within the theoretical framework of $D\rightarrow4$, the corresponding static spherically symmetric charged BH metric can be written as\cite{Fernandes:2020rpa}
\begin{equation}\label{eq3}
ds^2 = -f(r) dt^2 + f(r)^{-1} dr^2 + r^2 d\theta^2 + r^2 \sin^2\theta d\phi^2,
\end{equation}
\begin{equation}\label{eq4}
f(r) = 1 + \frac{r^2}{2\alpha} \left[ 1 \pm \sqrt{1 + 4\alpha \left( \frac{2M}{r^3} - \frac{Q^2}{r^4} \right)} \right].
\end{equation}
Where $M$ and $Q$ denote the mass and charge of the BH, respectively.
Unless otherwise specified, all variables in the following discussion are normalized by the black hole mass $M$. Accordingly, $r$, $Q$, $\alpha$, and $L$ denote the dimensionless quantities $r/M$, $Q/M$, $\alpha/M^2$, and $L/M$, respectively.

It should be pointed out that although the $D \to 4$ limiting scheme from which this BH solution originally arises has mathematical ambiguities at the action level, subsequent studies have shown that through various regularization methods and certain scalar-tensor theories, static spherically symmetric BH solutions consistent in form with Eq. (\ref{eq4}) can be obtained within well-defined four-dimensional theoretical frameworks method\cite{Lu:2020iav,Mann:1992ar,Kobayashi:2020wqy}. Therefore, investigating the particle dynamics and gravitational wave radiation properties in this charged BH background helps to understand the effects of Gauss-Bonnet corrections on strong-field spacetime structure, and may provide potential theoretical references for future space-based gravitational wave detection, thereby offering some basis for exploring the observational effects of higher-order curvature modified gravity theories.

The above BH solution has two branches. When $Q=0$ and $\alpha=0$, the negative branch reduces to the Schwarzschild spacetime; when $Q=0$, both branches reduce to the uncharged 4D-EGB BH in Ref.\cite{Glavan:2019inb}; in the limit of $\alpha \to 0$ and $r \to \infty$, only the negative branch can reduce to the Reissner-Nordström black hole(RNBH). Since the negative branch exhibits reasonable asymptotic behavior in all physical limits, this paper only considers this branch solution.

\begin{figure}[h]
\centering
\includegraphics[width=0.45\textwidth]{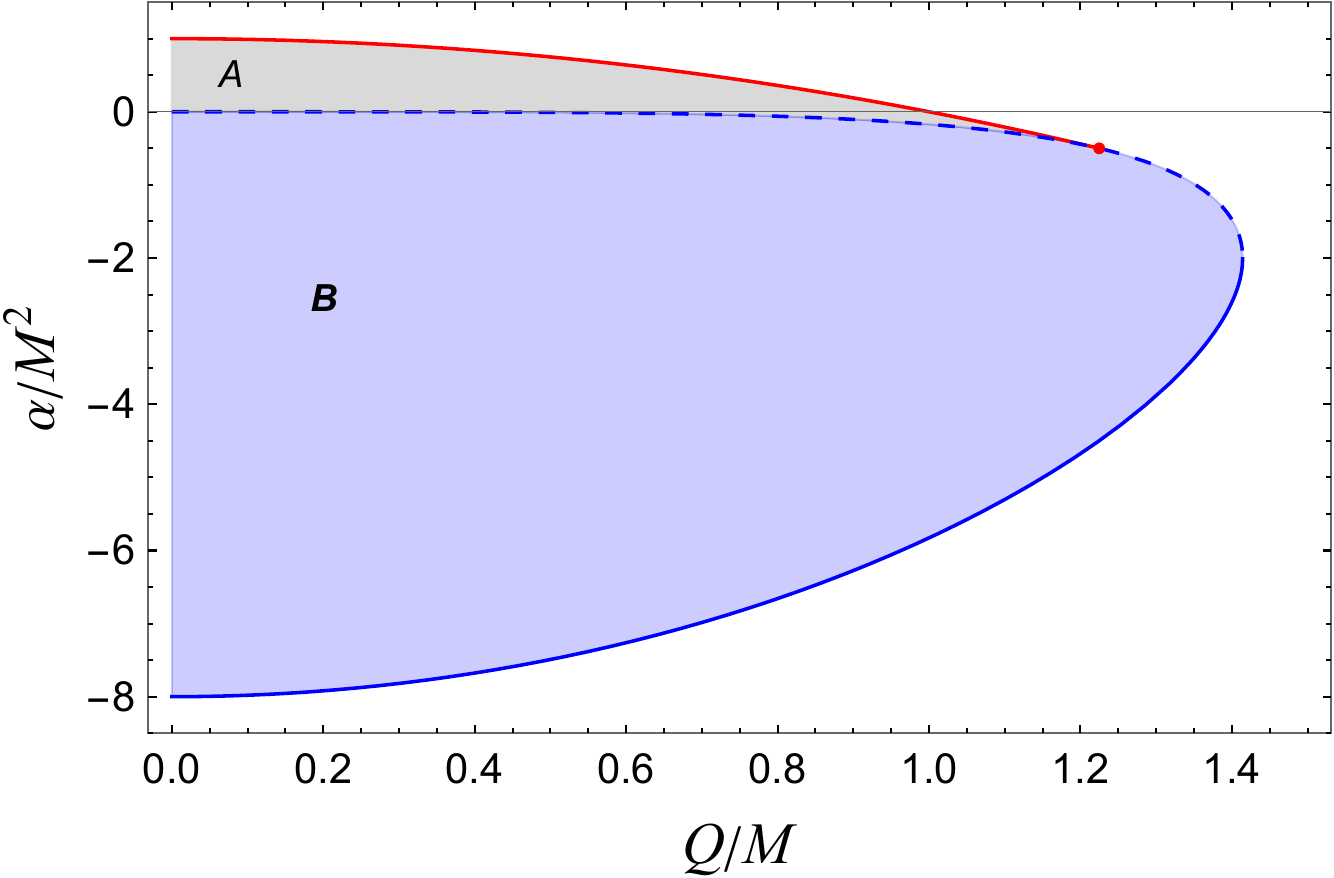}
\caption{
The allowed parameter space of charge $Q$ and coupling parameter $\alpha$ for the existence of BH.}
\label{a}
\end{figure}

The position of the BH event horizon is determined by $g^{rr}=0$, i.e., satisfying $f(r)=0$. Solving Eq. (\ref{eq4}) yields
\begin{equation}\label{eq5}
r_h = M \pm \sqrt{M^2 - Q^2 - \alpha}.
\end{equation}
Obviously, the BH event horizon exists when $M^2-Q^2-\alpha \geq 0$. Since this paper focuses on the physical properties outside the horizon, the Gauss-Bonnet coupling parameter $\alpha$ is allowed to take negative values \cite{Zhang:2020qam}. To clearly illustrate the parameter space for BH existence, we plot the $\alpha$ vs $Q$ parameter plane(see Fig. \ref{a}). As shown in the figure: within the parameter range $0<Q<\sqrt{3/2}$ and $-4+Q^2+2\sqrt{2}\sqrt{2-Q^2}<\alpha<1-Q^2$ (region A), the BH possesses a double-horizon structure; within the parameter range $0<Q<\sqrt{2}$ and $-4+Q^2-2\sqrt{2}\sqrt{2-Q^2}<\alpha<-4+Q^2+2\sqrt{2}\sqrt{2-Q^2}$ (region B), the BH has only a single horizon.

\subsection{Geodesics, MBO, and ISCO}
Since the charged BH is a static spherically symmetric solution in 4D-EGB gravity theory, without loss of generality, the particle motion can be confined to the equatorial plane (i.e., $\theta=\pi/2$, $\dot{\theta}=0$). In this case, considering the geodesic motion of a neutral test particle in the BH background, the Lagrangian is given by
\footnote{
In this work, the squared Lagrangian is adopted rather than the arc-length form $\mathcal{L} = \sqrt{-g_{\mu\nu}\dot{x}^\mu\dot{x}^\nu}$, which possesses time reparametrization invariance. Under affine parametrization, both formulations yield identical geodesic equations. Furthermore, the arc-length form degenerates to zero for null geodesics, rendering the variational principle ill-defined, whereas the squared form remains well-posed. The squared Lagrangian therefore constitutes the standard approach for treating both timelike and null geodesics in a unified framework (see, e.g., \cite{Haroon:2025rzx,Errehymy:2025pfr,Khan:2025qsy}).}

\begin{equation}
	\mathcal{L}=\frac{1}{2}g_{\mu\nu}\dot{x}^\mu\dot{x}^\nu=\frac{1}{2}\left(-f(r)\dot{t}^2+f(r)^{-1}\dot{r}^2+r^2\dot{\phi}^2\right),
\end{equation}
where the dot denotes differentiation with respect to the affine parameter. Since the Lagrangian does not explicitly depend on $t$ and $\phi$, the generalized momenta $p_t$ and $p_\phi$ are conserved, corresponding to the energy $E=-f(r)\dot{t}$ and angular momentum $L=r^2\dot{\phi}$, respectively.

Using the four-velocity normalization condition $g_{\mu\nu}\dot{x}^\mu\dot{x}^\nu=\varepsilon$, the radial equation of motion for the particle can be obtained as
\begin{equation}\label{eq7}
	\dot{r}^2=E^2-V_{\mathrm{eff}},
\end{equation}
where the effective potential is defined as
\begin{equation}\label{veff}
	V_{\mathrm{eff}}=f(r)\left(\frac{L^2}{r^2}-\varepsilon\right).
\end{equation}
The parameter $\varepsilon=-1$ corresponds to timelike geodesics, and $\varepsilon=0$ corresponds to null geodesics.

\begin{figure}[h]
\centering
\includegraphics[width=0.45\textwidth]{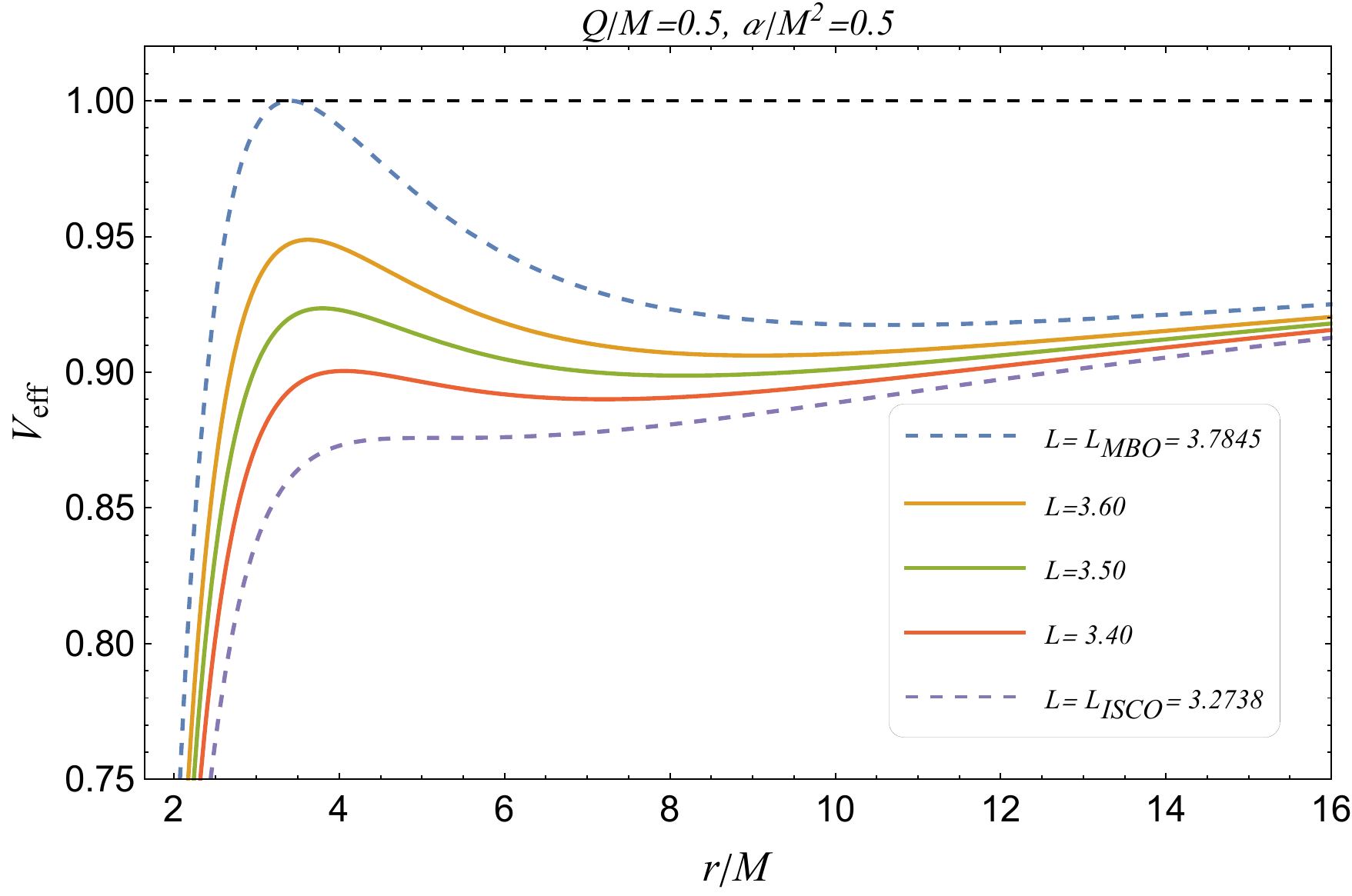}
\caption{Effective potential for different $L$.}
\label{b}
\end{figure}

To investigate the properties of the effective potential for particles, we present in Fig.\ref{b} the relationship between the effective potential $V_{\mathrm{eff}}$ and $r$ for timelike geodesics. Fig.\ref{b} shows that, under fixed $Q$ and $\alpha$, the extremum positions of the effective potential vary with the angular momentum $L$. When the angular momentum $L=L_{\mathrm{ISCO}}=3.2738$, corresponding to the purple dashed line in the figure, the two extremum points of the effective potential converge to a single point, and the particle is in a stable orbit; when the angular momentum $L=L_{\mathrm{MBO}}=3.7845$, corresponding to the blue dashed line in the figure, the effective potential reaches its maximum value with $V_{\mathrm{eff}}=E=1$, and the particle is in an unstable orbit. Furthermore, as $r\rightarrow\infty$, the effective potential $V_{\mathrm{eff}}\rightarrow1$. Therefore, combined with Eq. (\ref{eq7}), it can be deduced that bound orbits exist for particles when $E\leq1$.

\begin{figure*}[]
\includegraphics[width=0.9\textwidth]{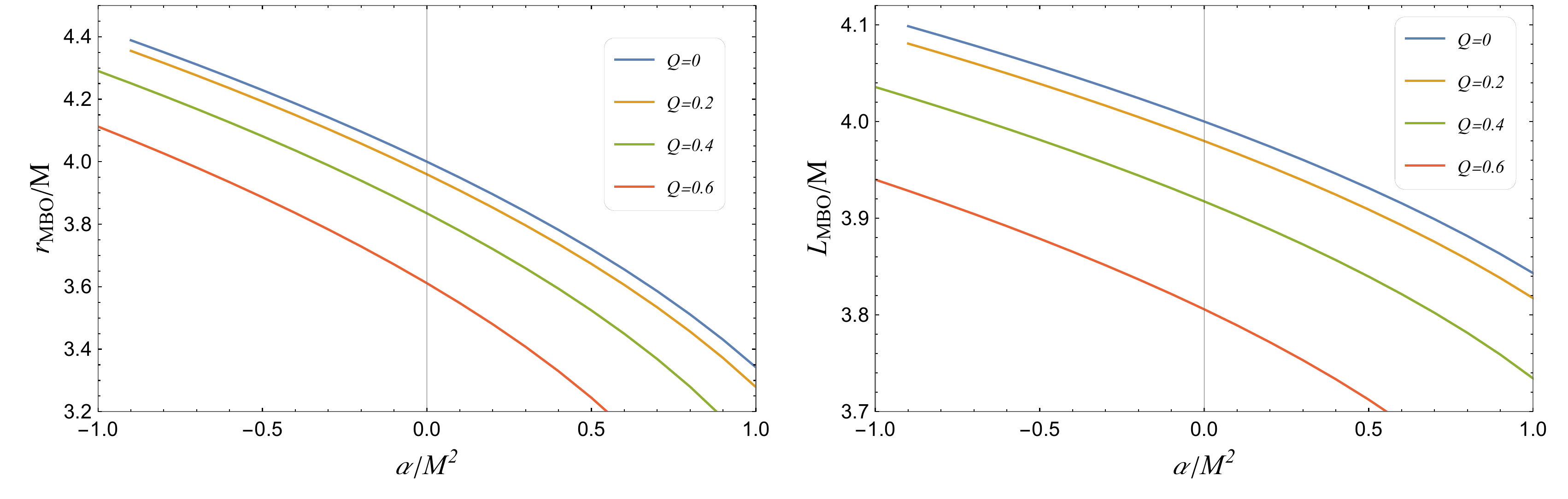}
\caption{Relationship between the marginally bound orbit radius $r_{MBO}$ and angular momentum $L_{MBO}$ with the coupling parameter $\alpha$.}
\label{c}
\end{figure*}

Next, we will explore the properties of particle orbits at the extrema of the effective potential, as shown by the blue dashed line in Fig. \ref{b}. We first consider the MBO around the BH. The MBO is an important circular orbit for studying BH physical information, and particles located on this orbit satisfy the following conditions
\begin{equation}\label{21}
V_{eff} = 1, \quad \partial_r V_{eff} = 0.
\end{equation}
By numerically solving the above equations, we can plot the relationship between the MBO radius $r_{\mathrm{MBO}}$ and the coupling parameter $\alpha$, as well as the relationship between the orbital angular momentum $L_{\mathrm{MBO}}$ and the coupling parameter $\alpha$, as shown in Fig.  \ref{c}. From Fig. \ref{c}, it can be observed that, for different values of charge $Q$, both $r_{\mathrm{MBO}}$ and $L_{\mathrm{MBO}}$ exhibit a clear monotonic decreasing trend as $\alpha$ increases. Specifically, in the parameter region with $\alpha>0$, compared to the Schwarzschild BH ($Q=0$, $\alpha=0$) and RNBH ($Q\neq0$, $\alpha=0$), BH in 4D-EGB theory have smaller $r_{\mathrm{MBO}}$ and $L_{\mathrm{MBO}}$; while for $\alpha<0$, they exhibit the opposite behavior. The sensitivity of these characteristic parameters to $\alpha$ and $Q$ provides potential possibilities for distinguishing BHs under different gravitational theory frameworks.

\begin{figure*}[]
\includegraphics[width=1\textwidth]{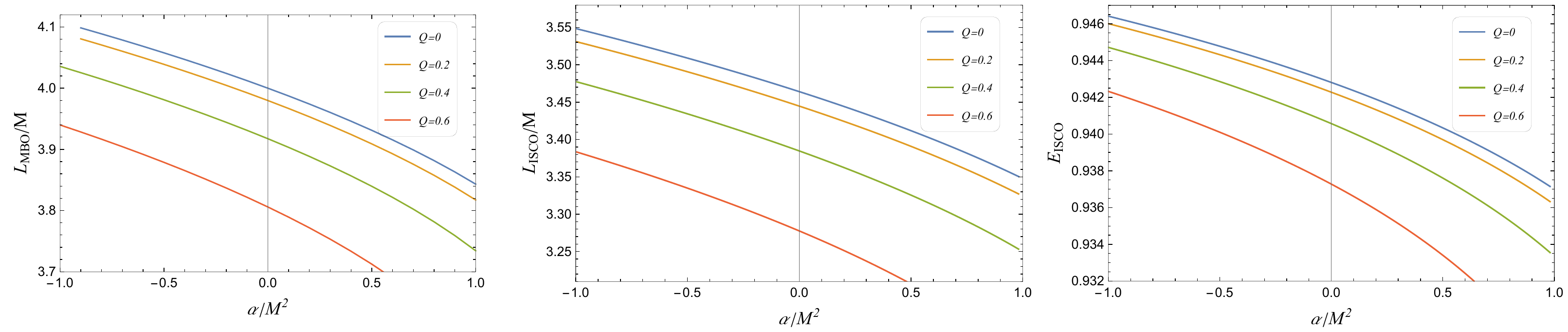}
\caption{Relationship between the innermost stable circular orbit radius $r_{ISCO}$, angular momentum $L_{ISCO}$, and energy $E_{ISCO}$ with the coupling parameter $\alpha$.}
\label{d}
\end{figure*}

The ISCO  refers to the critical minimum radius at which a test particle can maintain stable circular motion in BH spacetime. Particles on this orbit satisfy the following conditions
\begin{equation}\label{22}
\dot{r} = 0, \quad \partial_r V_{eff} = 0, \quad \partial_r^2 V_{eff} = 0.
\end{equation}
By numerically solving the above equations, the relationship between ISCO-related orbital parameters and the coupling parameter $\alpha$ can be plotted, as shown in Fig. \ref{d}. From Fig. \ref{d}, it is found that as the coupling parameter $\alpha$ increases, $r_{\mathrm{ISCO}}$, $L_{\mathrm{ISCO}}$, and $E_{\mathrm{ISCO}}$ all exhibit a gradually decreasing trend. In addition, a larger charge $Q$ leads to smaller $r_{\mathrm{ISCO}}$, $L_{\mathrm{ISCO}}$, and $E_{\mathrm{ISCO}}$. Compared to the Schwarzschild BH (i.e., $Q=0$, $\alpha=0$), when $\alpha$ is negative, particles can maintain stable circular motion on orbits farther from the BH; while when $\alpha$ is positive, particles can maintain stable circular motion on orbits closer to the BH. Furthermore, the RNBH ($Q\neq0$, $\alpha=0$) has smaller $r_{\mathrm{ISCO}}$, $L_{\mathrm{ISCO}}$, and $E_{\mathrm{ISCO}}$ compared to the Schwarzschild BH ($Q=0$, $\alpha=0$), which is consistent with the conclusions in related literature\cite{Mustafa:2024kjy,Zhao:2024exh}.

\begin{figure}[h]
\centering
\includegraphics[width=0.45\textwidth]{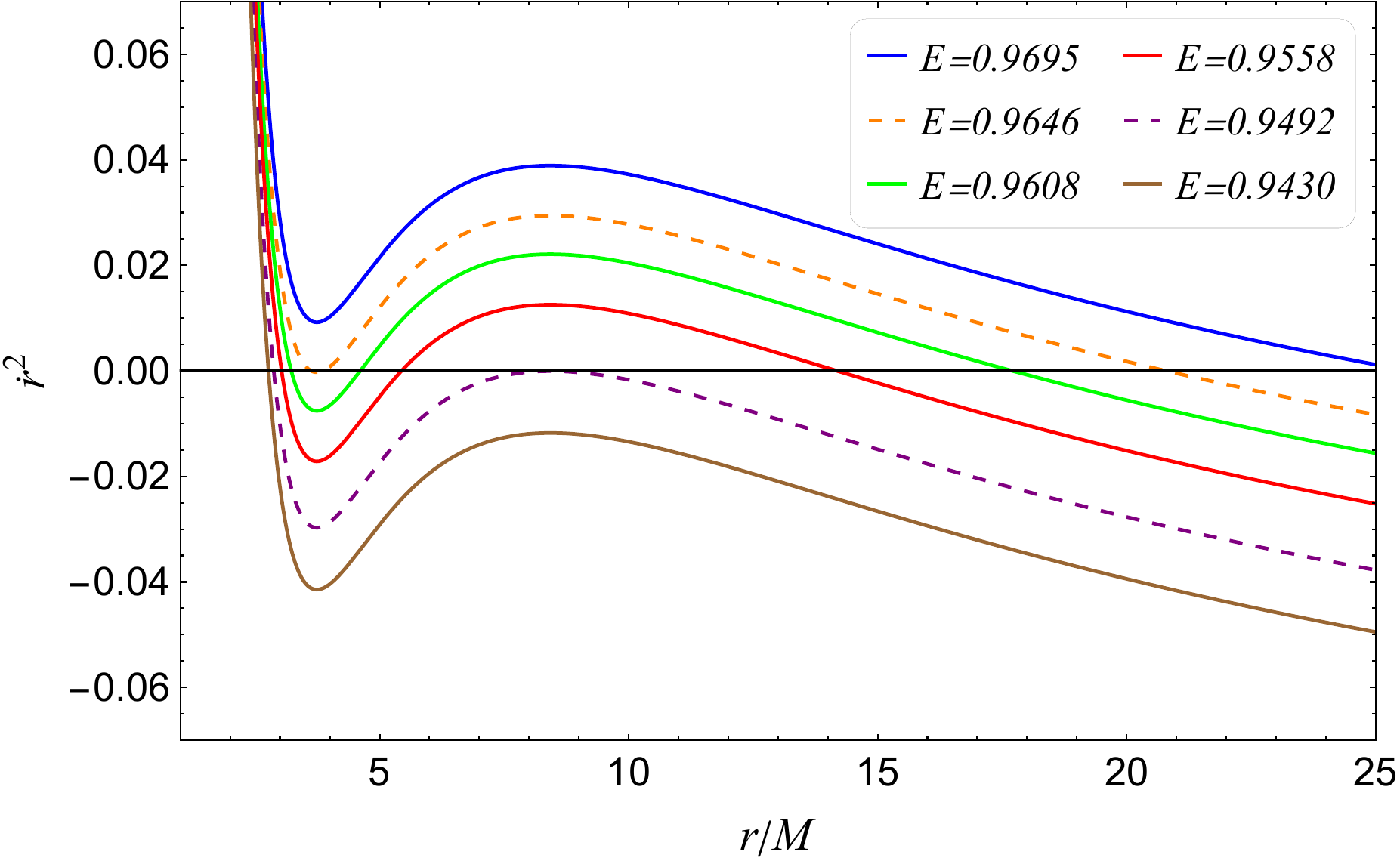}
\caption{Relationship between $\dot{r}^2$ and $r$, where $Q/M=0.5$, $\alpha/M^2=0.5$, and $L=0.5(L_{MBO}+L_{ISCO})$.}
\label{e}
\end{figure}

\begin{figure*}[]
\includegraphics[width=0.9\textwidth]{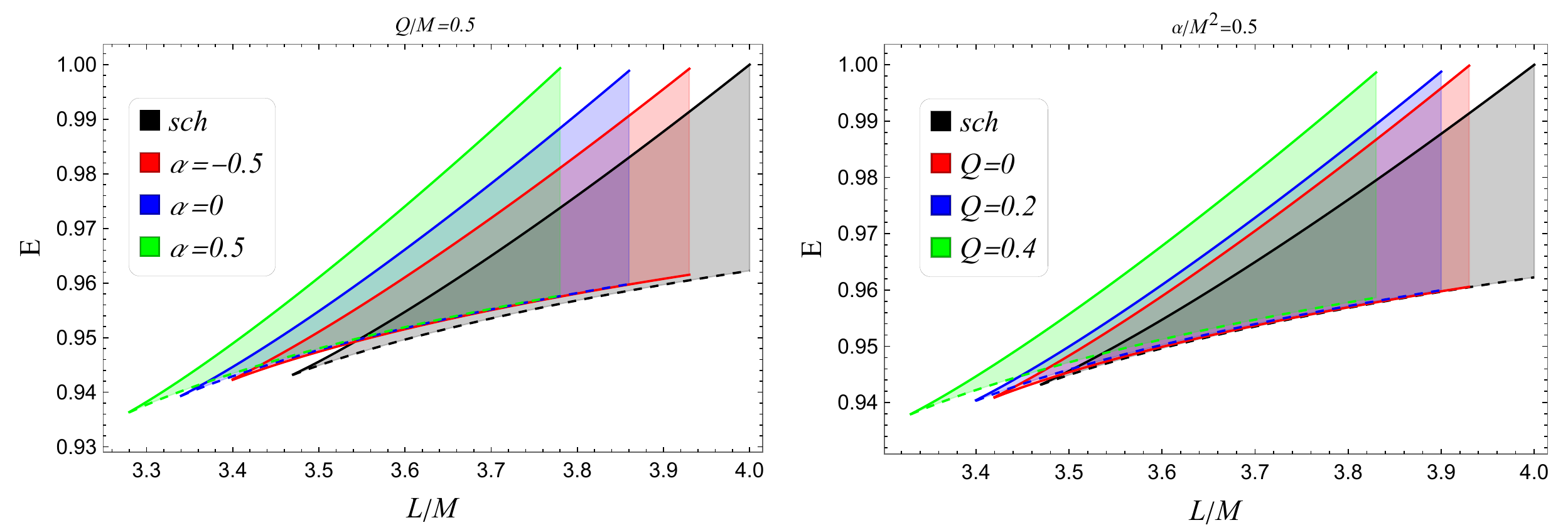}
\caption{Allowed parameter space ($E, L$) for bound orbits of the particle around a charged BH in 4D-EGB theory.}
\label{f}
\end{figure*}

According to the preceding analysis, although $E \leq 1$ is a necessary condition for a particle to be confined within a finite radial region, the existence of bound orbits is still subject to the joint constraints of energy $E$ and angular momentum $L$. This means that, under a given angular momentum condition, only particles within a specific energy interval can satisfy the conditions for maintaining bound orbit motion. As shown in Fig. \ref{e}, taking the case with fixed parameters $Q=0.5$, $\alpha=0.5$, and angular momentum $L=0.5(L_{\mathrm{MBO}}+L_{\mathrm{ISCO}})$ as an example, the energy interval allowing bound orbits to exist is $(0.9492, 0.9646)$, i.e., only orbits between the purple dashed line and the orange dashed line represent bound orbits. 
If the energy exceeds this range, no bound orbits will exist in the system.
When $E > 0.9646$, the particle has sufficient energy to overcome the effective potential barrier, thereby spiraling toward the BH from a distance and eventually being captured; while when $E < 0.9492$, since the energy is below the minimum energy level required to maintain orbital balance, the particle cannot form a stable bound state, manifesting as a direct spiral plunge toward the BH center. Furthermore, Fig. \ref{f} provides a more intuitive representation, where the black shaded region represents the allowed parameter interval $(E, L)$ for bound orbits in the Schwarzschild BH. From Fig. \ref{f}, it is found that when $Q=0.5$ is fixed, as the coupling parameter $\alpha$ increases, the triangular shaded region gradually shifts to the left; while when $\alpha=0.5$ is fixed, as the charge $Q$ increases, the triangular shaded region also shifts to the left. These results indicate that the presence of charge $Q$ and coupling parameter $\alpha$ significantly alters the allowed parameter region for bound orbits.

\section{Constraints on the coupling parameter and charge}\label{3.0}
In Section \ref{2.1}, we have provided the ranges of the coupling parameter $\alpha$ and charge $Q$ based on the fact that the BH possesses an event horizon, as shown in the shaded region in Fig. \ref{a}. Next, we will combine observational data from BH shadows and precession to impose more stringent physical constraints on $Q$ and $\alpha$, so as to make them more consistent with realistic physical models.

\subsection{Constraints from EHT}\label{3.1}

The Event Horizon Telescope (EHT) observations of $M87^{\ast }$ and $Sgr A^{\ast }$ provide a valid constraint on the parameters of the 4D-EGB gravity theory. Based on the observational data of $M87^{\ast }$ and $Sgr A^{\ast }$ reported in literature \cite{EventHorizonTelescope:2019dse,EventHorizonTelescope:2019ggy,EventHorizonTelescope:2019uob,EventHorizonTelescope:2022wkp,EventHorizonTelescope:2022wok}, the shadow sizes for $M87^{\ast }$ and $Sgr A^{\ast }$ at a $1\sigma$ confidence level are\cite{EventHorizonTelescope:2021dqv,Vagnozzi:2022moj,EventHorizonTelescope:2020qrl}
\begin{align}
4.313M \leq R_{sh}^{M87^{\ast }} \leq 6.079M,\label{23}\\
4.547M \leq R_{sh}^{Sgr A^{\ast }} \leq 5.222M.	\label{24}
\end{align}

Next, we investigate the shadow characteristics of charged BH in the 4D-EGB theoretical framework. To determine the observed radius of the BH shadow, the primary task is to determine the photon sphere radius. For null geodesics ($\varepsilon=0$), the effective potential (\ref{veff}) can be written as
\begin{equation}
	V_{\mathrm{eff}}^{ph}=f(r)\left(\frac{L^2}{r^2}\right).
\end{equation}
The photon sphere radius $r_{ph}$ corresponds to an unstable null circular orbit, with the conditions
\begin{equation}
	V_{\mathrm{eff}}^{ph}=E^2,\ \ \partial_r V_{\mathrm{eff}}^{ph}\bigg|_{r=r{ph}}=0,\ \ \partial_r^2 V_{\mathrm{eff}}^{ph}\bigg|_{r=r{ph}}<0.
\end{equation}
Solving the above equations yields
\begin{equation}
	r_{ph}f^\prime(r_{ph})-2f(r_{ph})=0.
\end{equation}
For a distant observer, the relationship between the BH shadow radius and the photon sphere radius is \cite{EventHorizonTelescope:2020qrl}
\begin{equation}
	R_s=\frac{r_{ph}}{\sqrt{f(r_{ph})}}.
\end{equation}

\begin{figure}[h]
\includegraphics[width=0.45\textwidth]{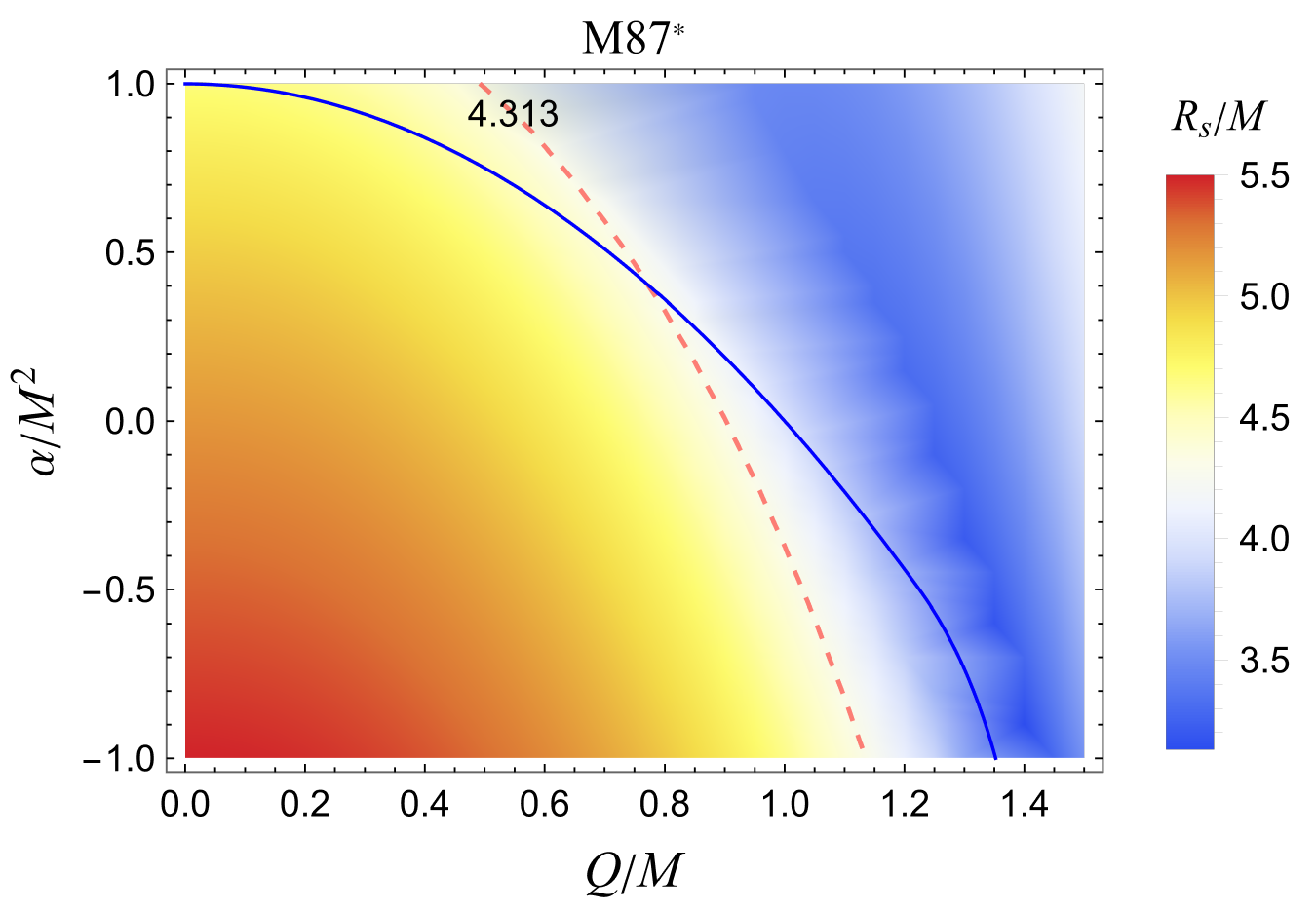}
\caption{Constraints on the  $\alpha$ and  $Q$ using the EHT observational data of M87*. The red dashed line represents the minimum shadow radius of M87*, and the blue curve represents the constraint curve based on BH existence.}
\label{g}
\end{figure}

\begin{figure}[h]
\includegraphics[width=0.45\textwidth]{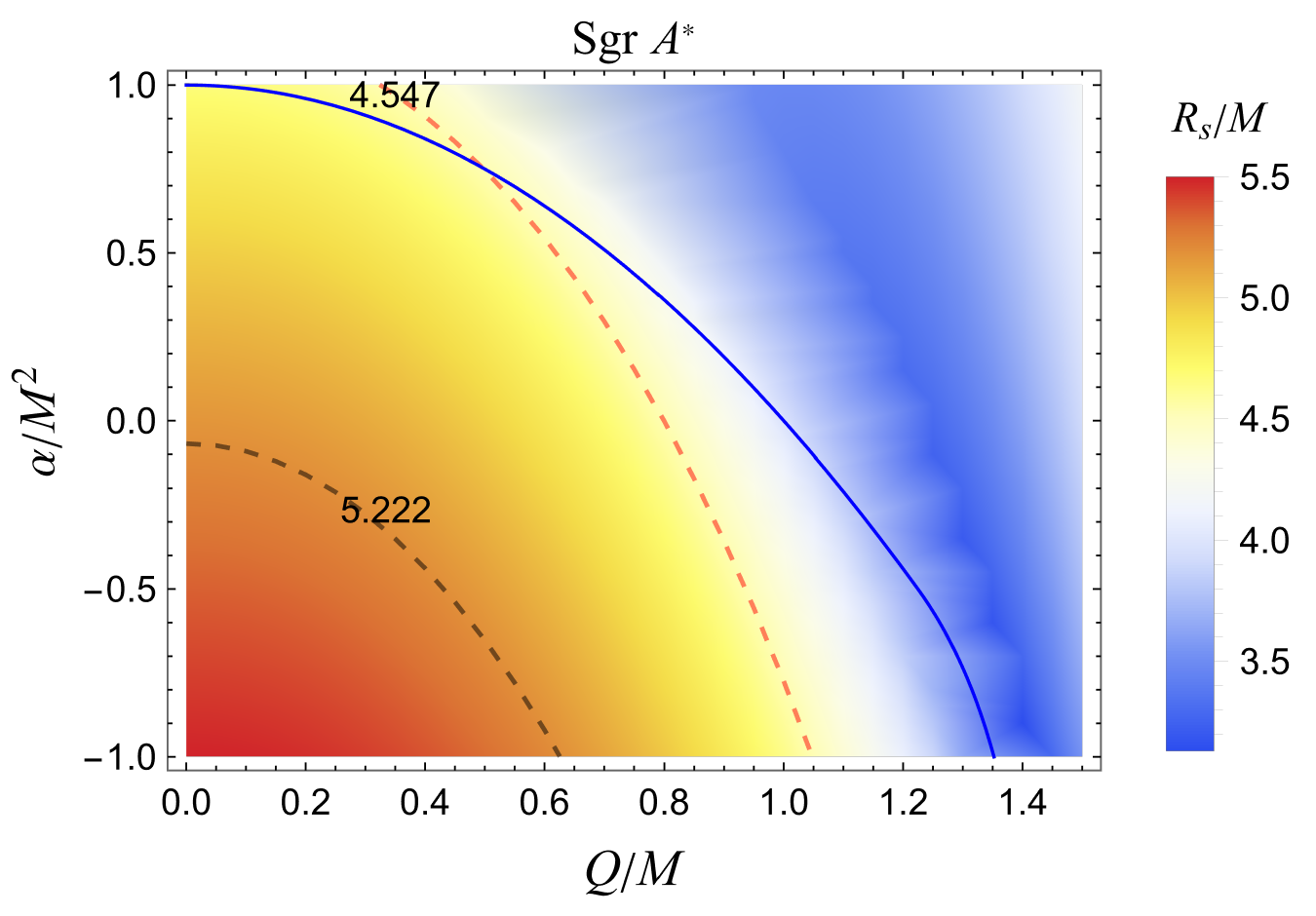}
\caption{Constraints on the $\alpha$ and  $Q$ using the EHT observational data of Sgr A*. The region between the red dashed line and the black dashed line represents the shadow range of Sgr A*, and the blue curve represents the constraint curve based on BH existence.}
\label{h}
\end{figure}

We consider the BH model in the 4D-EGB gravity theoretical framework as a theoretical candidate for M87* and Sgr A*, and use their observational data to constrain the model parameter space. As shown in Figs. \ref{g} and \ref{h}, since the observational upper limit of the shadow radius $R_s/M$ is 5.5, only the lower limit constraint region of the M87* shadow radius is presented in Fig. \ref{g}. From Fig. \ref{g}, it can be seen that the blue curve represents the boundary condition between the BH and the naked singularity, and the enclosed region represents the theoretical parameter space for BH existence. Compared to this theoretically allowed region, the EHT observational data of M87* imposes certain constraints on the charge parameter $Q$, but has a relatively weak constraining ability on the coupling parameter $\alpha$. Specifically, the observationally constrained region (the intersection of the red dashed line and the blue solid line) shows a significant contraction compared to the BH existence region (the region enclosed by the blue curve), indicating that the observational data further excludes some parameter combinations that theoretically allow the existence of an event horizon but are inconsistent with observations. Furthermore, according to the Sgr A* shadow radius constraint results shown in Fig. \ref{h}, it can be found that the EHT observational data imposes more stringent constraints on the 4D-EGB BH model parameters. Overall, compared to M87*, the observational data of Sgr A* provides stronger parameter constraining capability.

\subsection{Constraints from $S2$ star precession}\label{3.2}
Besides BH shadows, the orbital dynamical behavior of stars around the supermassive BH Sgr A* also provides a natural laboratory for testing gravitational theories. Based on the long-term monitoring of the S2 star orbit by the GRAVITY Collaboration, the fitting analysis range between the observed value of the S2 star perihelion precession $\Delta\omega_{S2}$ and the theoretical predicted value $\Delta\omega_{GR}$ within the GR framework is  \cite{GRAVITY:2020gka}
\begin{equation}\label{31}
f_{sp} = \frac{\Delta \omega_{S2}}{\Delta \omega_{GR}} = 1.10 \pm 0.19.
\end{equation}
Although this result is consistent with the predictions of GR within the error range, the uncertainties present in the observational data still leave potential physical space for extended gravity models such as 4D-EGB theory to deviate from the standard gravitational framework. Accordingly, by comparing the precession value predicted by 4D-EGB theory with the above observational interval, we can further constrain the allowed ranges of the model parameters $Q$ and $\alpha$.

According to the research method in literature \cite{Levin:2008mq}, by defining a parameter $q$ to describe and classify the orbital properties around a BH, the parameter $q$ is defined as
\begin{equation}\label{32}
q = \frac{\Delta \phi}{2\pi} - 1.
\end{equation}
Here, when the parameter $q$ is a rational number, it indicates that the particle's orbit around the BH is a periodic orbit, meaning the particle returns precisely to its starting position after one revolution (its properties will be detailed in the next section). When the parameter $q$ is an irrational number, it indicates that the particle's orbit is a precessing orbit, meaning there will be an orbital deviation between two adjacent orbits of the particle (this subsection studies the precession phenomenon), which will then produce an orbital precession angle $\Delta\omega$, where $\Delta\omega = \Delta\phi - 2\pi$. Here, $\Delta\phi$ is consistent with that shown in Eq.  \eqref{32}, representing the angular displacement generated when a particle moves back and forth between the two turning points of a bound orbit (as shown in Fig. \ref{e}), and the corresponding expression is
\begin{equation}\label{33}
\Delta \phi = \oint d\phi = 2 \int_{r_1}^{r_2} \frac{d\phi}{dr} dr,
\end{equation}
where $r_1$ and $r_2$ represent the two turning points.

To facilitate the calculation of the orbital precession angle $\Delta\omega_C$ for particle precession around a charged BH in the 4D-EGB gravity theory, we use the formula for the trajectory of particle around a BH, which is expressed as\cite{Chandrasekhar:1984siy,Straumann:2013spu}
\begin{equation}\label{34}
r = \frac{a(1 - e^2)}{1 + e \cos \psi}.
\end{equation}
When $\psi=0$, the pericenter is $r_a=a(1-e)$; when $\psi=\pi$, the apocenter is $r_b=a(1+e)$. Here $\psi$ denotes the relativistic true anomaly, $a$ denotes the orbital semi-major axis, and $e$ denotes the eccentricity.

Combining Eqs. (\ref{33}),(\ref{34}) and the orbital differential relation derived from the conservation of generalized momentum, we obtain
\begin{equation}\label{35}
\Delta \phi = 2 \int_{r_a}^{r_b} \frac{d\phi}{dr} dr = 2 \int_{0}^{\pi} \frac{d\phi}{dr} \frac{dr}{d\psi} d\psi,
\end{equation}
where
\begin{equation}\label{36}
\frac{d\phi}{d\psi} =\frac{a L e (1 - e^2) \sin \psi}{r^2 (1 + e \cos \psi)^2 \sqrt{E^2 - f(r) \left( 1 + \frac{L^2}{r^2} \right)}}.
\end{equation}
In the above equation, $E^2$ and $L^2$ are determined by the turning point conditions $\dot{r}=0$ at the pericenter $r_a$ and apocenter $r_b$.

Expanding Eq. (\ref{36}) in the weak-field parameter $M/a$ ($M/a \ll 1$) and integrating term by term with respect to $\psi$, we obtain the degree of deviation between 4D-EGB gravity and GR, whose expression is given by
\begin{equation}\label{40}
f_{sp} = \frac{\Delta \omega_{C}}{\Delta \omega_{GR}},
\end{equation}
where
\begin{widetext}
	\begin{align}\label{41}
\Delta \omega_{C} =&\frac{  \pi \left(6 - Q^2\right)  M }{ \left(1 - e^2\right) a} 
- \frac{ \pi \left(-108 + 48 Q^2 + Q^4 + 2 e^2 \left(-3 + Q^2\right)\right)   M^2 }{4  \left(1 - e\right)^2 \left(1 + e\right)^2 a^2} \notag
\\
&- \frac {\pi \left(756 Q^2 - 62 Q^4 + Q^6 + 24 \left(-45 + 8 \alpha\right) + 2 e^2 \left(-90 + 57 Q^2 + Q^4 + 24 \alpha\right)\right)M^3 }{8  \left(1 - e^2\right)^3 a^3}+ \mathcal{O}\left(\frac{M^4}{a^4}\right),
\end{align}
\end{widetext}
with $\Delta \omega_{GR} = 6\pi M / [a(1 - e^2)]$ being the standard GR result. Note that the weak-field condition $M/a \ll 1$ is well satisfied for the S2 star orbiting Sgr A*, justifying the validity of the above expansion.

\begin{figure}[h]
\centering
\includegraphics[width=0.45\textwidth]{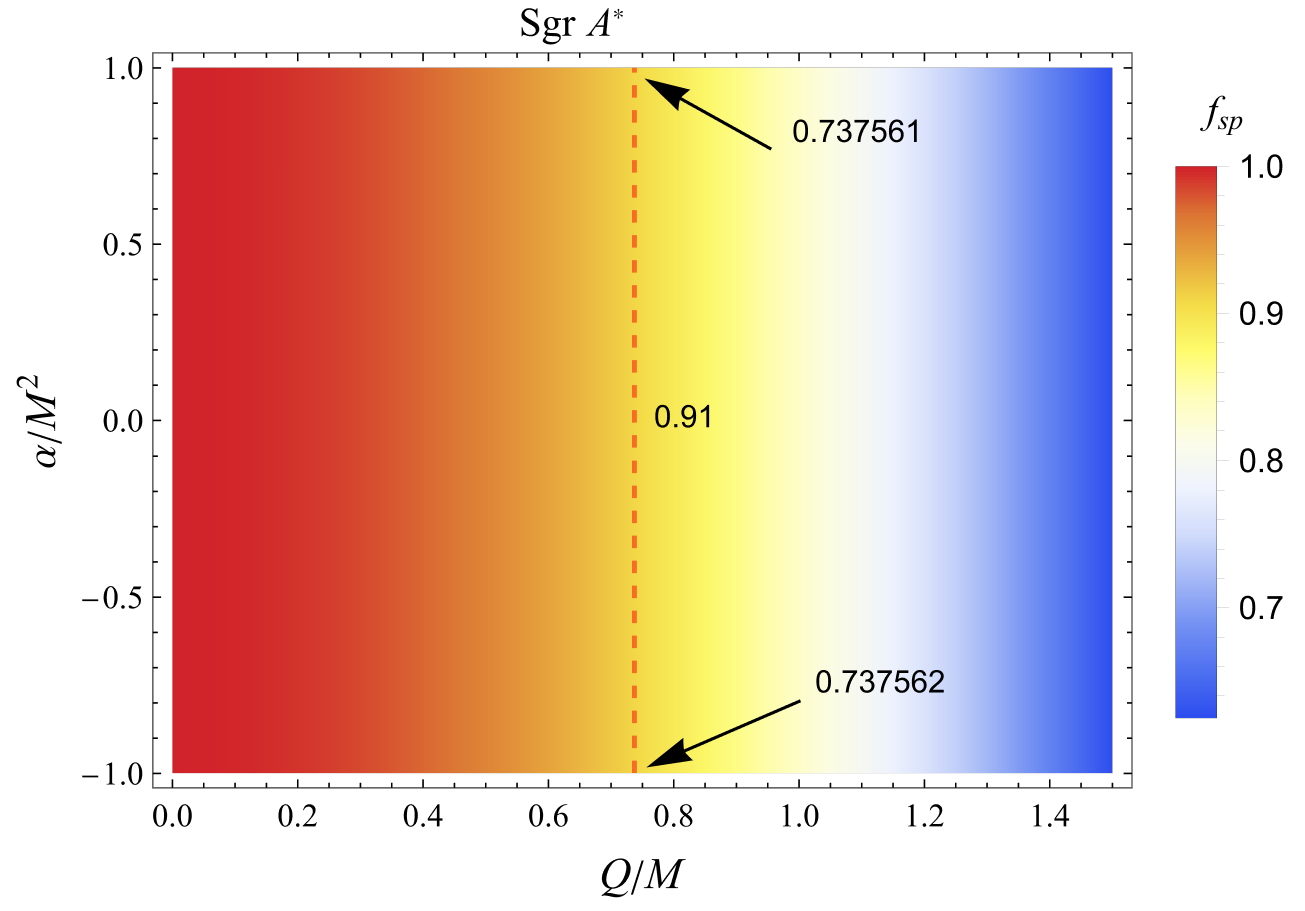}
\caption{Using $S2$ star's precession observational data to constrain parameter $\alpha$ and charge $Q$.}
\label{i}
\end{figure}

Now, based on the observational data of the S2 star precession \cite{GRAVITY:2020gka}, we further obtain the constraint ranges of the parameter $\alpha$ and charge $Q$ for the charged BH in 4D-EGB gravity theory, as shown in Fig. \ref{i}. From Fig. \ref{i}, it is found that the observational data of the S2 star precession only imposes stringent constraints on the charge $Q$, with the constraint range approximately being $(0, 0.738)$, while no significant constraints are formed on the parameter $\alpha$.
This can be intuitively understood from Eq. (\ref{41}): the contribution of the charge $Q$ to orbital precession emerges already at the leading order $\mathcal{O}(M/a)$, whereas the parameter $\alpha$ first appears only in the third-order correction term $\mathcal{O}(M^3/a^3)$. Since the S2 star satisfies the weak-field condition $M/a \ll 1$, the correction term dependent on $\alpha$ is numerically negligible, rendering this observable effectively incapable of placing stringent constraints on $\alpha$.
This further implies that to impose valid constraints on the parameter $\alpha$, investigations must be carried out in the strong-field regime; for instance, the detectability of $\alpha$ can be explored via the gravitational wave waveforms corresponding to periodic orbits around BHs.

\section{Periodic Orbits}\label{4.0}
In strong gravitational fields, periodic orbits of particles around BHs constitute an important class of bound orbits. They can intuitively characterize the spacetime geometry around BHs and theoretically predict the physical information of BHs. In this section, we will investigate the effects of energy $E$ and angular momentum $L$ on periodic orbits, and explore whether the properties of periodic orbits can serve as a basis for distinguishing Schwarzschild BH, RNBH, and charged BH in 4D-EGB gravity theory.

In this section of our study, we employ the periodic orbit classification method proposed by Janna Levin et al.\cite{Levin:2008mq} to investigate the properties of periodic orbits around BHs. This classification method uses three integers ($z, w, v$) to define a rational parameter $q$, where $q$ is expressed as
\begin{equation}\label{43}
q = \frac{\Delta \phi}{2\pi} - 1 = w + \frac{\nu}{z},
\end{equation}
here, $\Delta \phi$ is defined by Eq.  \eqref{35}, and $z$, $w$, and $v$ represent the zoom number, whirl number, and vertex number of the orbit, respectively. Combining the geodesic differential equation, the above equation can be written as
\begin{align}\label{44}
q = \frac{1}{\pi} \int_{r_1}^{r_2} \frac{L}{r^2 \sqrt{E^2 - f(r) \left( 1 + \frac{L^2}{r^2} \right)}} dr - 1.
\end{align}

\begin{figure}[]
\includegraphics[width=0.5\textwidth]{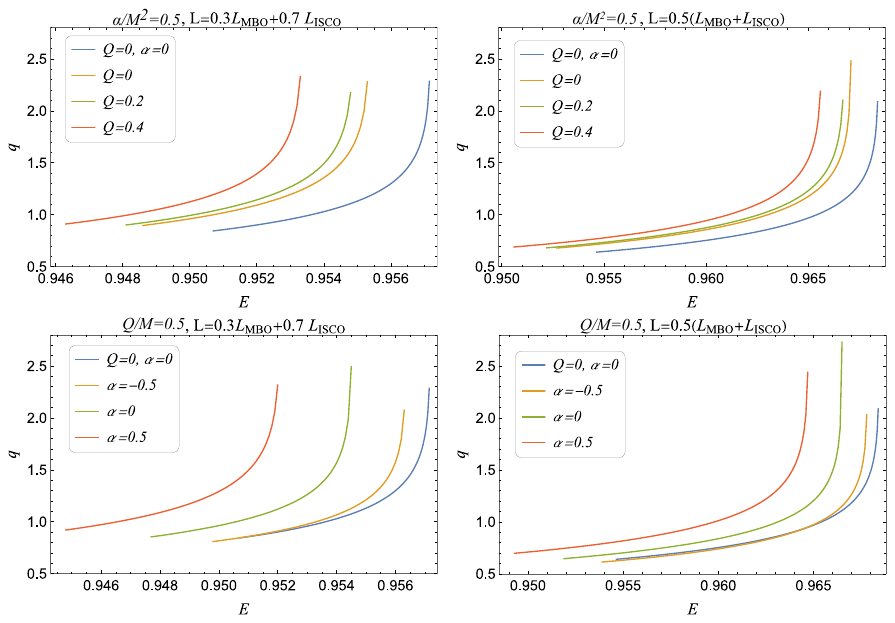}
\caption{The $q$ vs $E$ relationship diagram under different parameter values, where $Q/M = 0$ and $\alpha/M^2 = 0$ corresponds to the Schwarzschild case.}
\label{j}
\end{figure}

\begin{figure}[]
\includegraphics[width=0.5\textwidth]{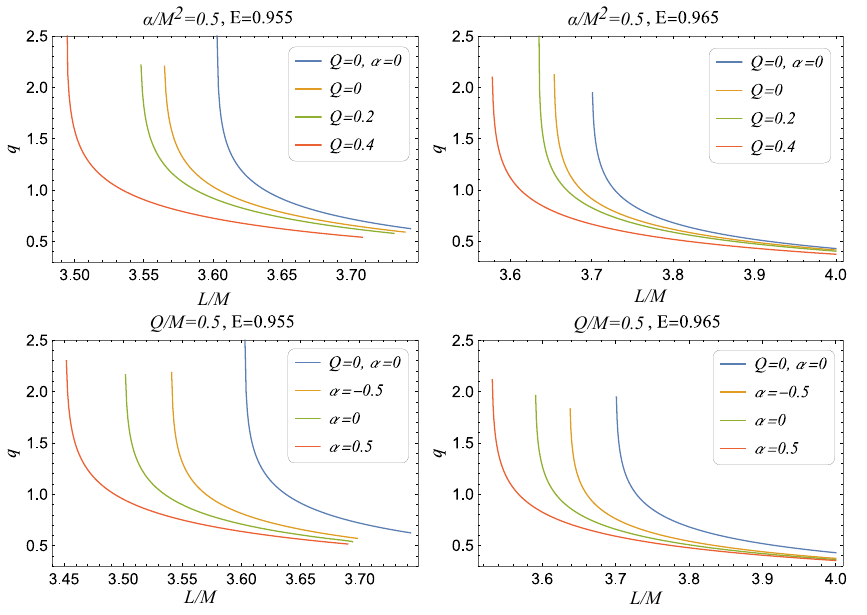}
\caption{The $q$ vs $L$ relationship diagram under different parameter values, where $Q/M = 0$ and $\alpha/M^2 = 0$ corresponds to the Schwarzschild case.}
\label{k}
\end{figure}

Next, by selecting appropriate values of the parameter $\alpha$ and charge $Q$, we plot the relationship diagrams of rational number $q$ vs $E$ and $q$ vs $L$ based on the above equations.
As can be seen from the analysis results in Figs.\ref{j} and \ref{k}, the parameter $\alpha$ and charge $Q$ significantly impact the energy and angular momentum of particle trajectories. Specifically, as parameter $\alpha$ and charge $Q$ increase, the energy and angular momentum of the evolving particle trajectories show a decreasing trend. Furthermore, it is noteworthy that under the same rational number $q$, both RNBH ($Q=0.5, \alpha=0$) and charged BH within the 4D-EGB gravity theory exhibit lower energy and angular momentum characteristics compared to the Schwarzschild BH case (represented by the blue solid line). These distinguishing features provide potential theoretical grounds for differentiating BH types using the properties of periodic orbits.

In order to plot the periodic orbital trajectories of the particle, we provide the numerical values for energy $E$ and angular momentum $L$ corresponding to different rational numbers $q$ in Table \ref{table1} and Table \ref{table2}. Based on the data analysis, it can be concluded that under the condition of a fixed parameter $q$, Schwarzschild BH exhibit the maximum values for both energy $E$ and angular momentum $L$. Furthermore, when the angular momentum $L$ is kept constant, the individual influence of parameter $\alpha$ or charge $Q$ on the particle's orbital energy value is not significant. However, when energy $E$ is kept constant, parameter $\alpha$ shows a more significant impact on the particle's orbital angular momentum value compared to charge $Q$. 

\begin{table*}[]
\centering
\caption{Energy $E$ for different parameter values, with fixed $L=0.5(L_{MBO}+L_{ISCO})$.}
\label{table1}
\begin{tabular}{w{l}{1.8cm}w{l}{1.8cm}w{l}{2.2cm}w{l}{2.2cm}w{l}{2.2cm}w{l}{2.2cm}w{l}{2.2cm}w{l}{2.2cm}}
\hline\hline
\rule{0pt}{12pt} $Q/M$&$\alpha /M^2$ &$E_{(1,1,0)}$ & $E_{(1,2,0)}$ & $E_{(2,1,1)}$ & $E_{(2,2,1)}$ & $E_{(3,1,2)}$ & $E_{(3,2,2)}$ \\
\hline
\rule{0pt}{12pt} 
0 & 0 & 0.965425 & 0.968383 & 0.968026 & 0.968434 & 0.968225 & 0.968438 \\
\rule{0pt}{12pt} 
0.5& -0.5 &  0.965309 & 0.967795 & 0.967518 & 0.967832 & 0.967675 & 0.967835  \\
\rule{0pt}{12pt} 
0.5&  0 & 0.963135 & 0.966433 & 0.966025 & 0.966493 & 0.966251 & 0.966498   \\
\rule{0pt}{12pt} 
 0.5& 0.5 &0.959805 & 0.964585 & 0.963901 & 0.964706 & 0.964266 & 0.964718  \\
\rule{0pt}{12pt} 
 0&0.5 &0.963032 & 0.967012 & 0.966475 & 0.967101 & 0.966766 & 0.967109  \\
\rule{0pt}{12pt} 
0.2& 0.5 & 0.962585 & 0.966666 & 0.966111 & 0.966759 & 0.966411 & 0.966767 \\
\rule{0pt}{12pt} 
0.4& 0.5 & 0.961097 & 0.965538 & 0.964918 & 0.965645 & 0.965251 & 0.965655 \\
\hline\hline
\end{tabular}
\end{table*}

\begin{table*}[]
\centering
\caption{Angular momentum $L$ for different parameter values, with fixed $E=0.965$.}
\label{table2}
\begin{tabular}{w{l}{1.8cm}w{l}{1.8cm}w{l}{2.2cm}w{l}{2.2cm}w{l}{2.2cm}w{l}{2.2cm}w{l}{2.2cm}w{l}{2.2cm}}
\hline\hline
\rule{0pt}{12pt} $Q/M$&$\alpha /M^2$ &$L_{(1,1,0)}$ & $L_{(1,2,0)}$ & $L_{(2,1,1)}$ & $L_{(2,2,1)}$ & $L_{(3,1,2)}$ & $L_{(3,2,2)}$ \\
\hline
\rule{0pt}{12pt} 
0 & 0 & 3.72830 & 3.70086 & 3.70441 & 3.70031 & 3.70246 & 3.70027  \\
\rule{0pt}{12pt} 
0.5& -0.5 & 3.66054 & 3.63758 & 3.64032 & 3.63720 & 3.63879 & 3.63717  \\
\rule{0pt}{12pt} 
0.5&  0 &3.61836 & 3.59090 & 3.59445 & 3.59035 & 3.59250 & 3.59030  \\
\rule{0pt}{12pt} 
 0.5& 0.5 &3.56772 & 3.53237 & 3.53757 & 3.53144 & 3.53480 & 3.53135  \\
\rule{0pt}{12pt} 
 0&0.5 &3.68752 & 3.65435 & 3.65906 & 3.65354 & 3.65653 & 3.65346  \\
\rule{0pt}{12pt} 
0.2& 0.5 & 3.66934 & 3.63594 & 3.64070 & 3.63512 & 3.63814 & 3.63504 \\
\rule{0pt}{12pt} 
0.4& 0.5 &3.61265 & 3.57831 & 3.58328 & 3.57744 & 3.58062 & 3.57735\\
\hline\hline
\end{tabular}
\end{table*}

\begin{figure}[]
\includegraphics[width=0.5\textwidth]{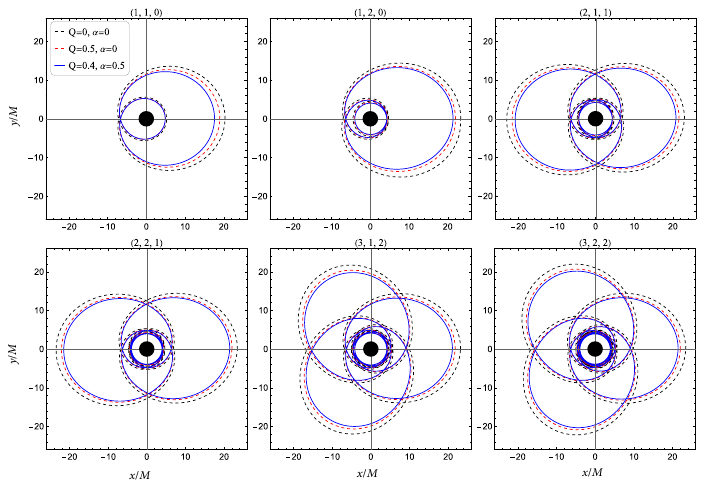}
\caption{Periodic trajectories of the particle under the condition of a fixed orbital angular momentum $L=0.5(L_{MBO}+L_{ISCO})$.}
\label{l}
\end{figure}

\begin{figure}[]
\includegraphics[width=0.5\textwidth]{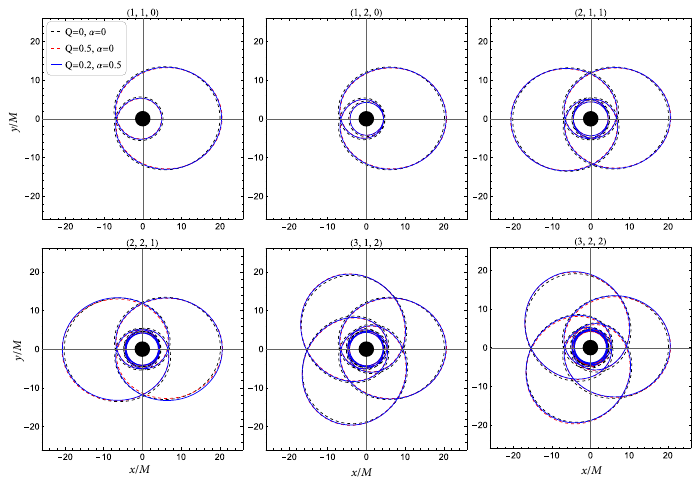}
\caption{Periodic trajectories of the particle under the condition of a fixed orbital energy $E=0.965$.}
\label{m}
\end{figure}

Based on the data in Tables \ref{table1} and \ref{table2}, we display in Figs.\ref{l} and \ref{m} the periodic trajectories of particles around Schwarzschild BH (black dashed lines), RNBH (red dashed lines), and charged BH in 4D-EGB gravity theory (blue solid lines).
 From these trajectories, it is easy to observe that: different rational numbers $q$ lead to more complex particle trajectories, mainly manifested in that a larger $z$ corresponds to a greater number of leaves, and a larger $w$ increases the number of rotations of the particle around the center; in 4D-EGB theory, the modifications to the spacetime structure by the coupling constant $\alpha>0$ and charge $Q$ allow stable periodic orbits of particles to exist in regions closer to the BH, while for $\alpha<0$, stable periodic orbits exist in regions farther from the BH (see the left panels of Figs.\ref{n} and \ref{o}); the differences in geometric morphology among the periodic orbits of the three types of BHs not only help to explore BH properties and distinguish different BH models, but also provide a possible avenue for testing 4D-EGB gravity theory through EMRI gravitational wave observations.

\section{Gravitational wave radiation of periodic orbits}\label{5.0}
This section will explore the gravitational wave radiation characteristics of periodic orbits in EMRIs. Such systems are formed by stellar-mass compact objects spiraling around supermassive BHs, where periodic orbits are transitional orbits in the inspiral evolution process. Since the timescale of orbital parameter changes caused by gravitational wave radiation is much longer than a single orbital period, and the energy and angular momentum loss within a single period is negligible compared to the total system energy and angular momentum, the adiabatic approximation can be applied to each instantaneous periodic orbit, i.e., the energy $E$ and angular momentum $L$ of the orbit are treated as constants during calculations.

We employ the Kludge waveform scheme \cite{Babak:2006uv} to calculate the gravitational wave radiation from periodic orbits around charged BH in 4D-EGB gravity theory, i.e., based on the discussion of periodic orbits in the previous section, we use the quadrupole approximation to compute the gravitational waveforms generated by periodic orbits. The quadrupole moment formula is given by \cite{Will:2016sgx,Liang:2022gdk,Maselli:2021men}
\begin{equation}\label{45}
h_{ij} = \frac{4 \eta M}{D_L} \left( v_i v_j - \frac{m}{r} n_i n_j \right),
\end{equation}
where $M$ represents the mass of the supermassive BH, $m$ is the particle mass, $D_L$ is the luminosity distance of the EMRI system, $\eta=\frac{Mm}{(M+m)^2}$ is the symmetric mass ratio, $v$ denotes the relative velocity of the particle, and $n$ is the unit vector in the radial direction.

In the transverse-traceless (TT) gauge, the plus polarization component $h_+$ and cross polarization component $h_\times$ can be obtained from Eq. \eqref{45} as\cite{Will:2016sgx}
\begin{align}
h_+ &= -\frac{2\eta M^2}{D_L r} (1 + \cos^2 \iota) \cos(2\phi + 2\xi),\label{46}\\
h_{\times} &= -\frac{4\eta M^2}{D_L r} \cos \iota \sin(2\phi + 2\xi).\label{47}
\end{align}
In the above equations, $\iota$ is the orbital inclination angle; $\xi$ represents the latitude angle, used to determine the projection of the particle's position on its orbit; $\phi$ is the phase angle, characterizing the time evolution parameter of the particle in periodic motion.

\begin{figure*}[]
\includegraphics[width=0.9\textwidth]{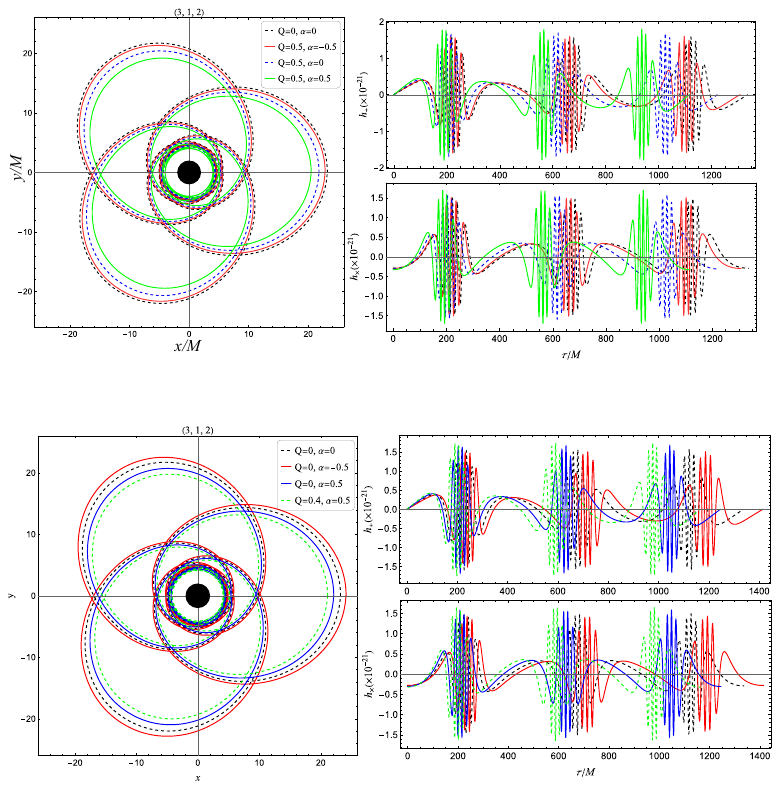}
\caption{The left panel shows the $(3,1,2)$ periodic orbital trajectories under different values of the coupling parameter $\alpha$; the right panel shows the corresponding gravitational waveforms of the periodic orbits.}
\label{n}
\end{figure*}

\begin{figure*}[]
\includegraphics[width=0.9\textwidth]{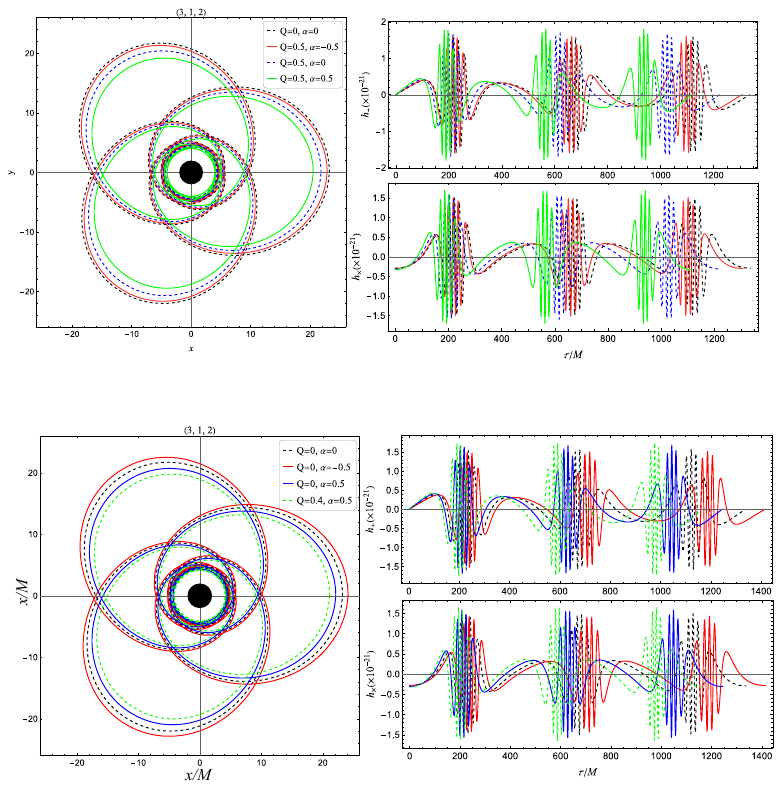}
\caption{The left panel shows the $(3,1,2)$ periodic orbital trajectories under different values of charge $Q$; the right panel shows the corresponding gravitational waveforms of the periodic orbits.}
\label{o}
\end{figure*}

To investigate the effects of parameters $\alpha$ and charge $Q$ on the gravitational wave signals of periodic orbits, we select several representative sets of $\alpha$ and $Q$ from the parameter space determined in Section \ref{3.0} by EHT observational constraints and the orbital precession of the S2 star, and adopt the $(3,1,2)$ periodic orbit as an example to calculate and plot the corresponding gravitational waveforms.
Here we set the EMRI system parameters as follows: $M=10^6 M_\odot$, $m=10M_\odot$, $\iota$ and $\xi$ are both $\frac{\pi}{4}$, $D_L=200$ Mpc. In the left panels of Figs. \ref{n} and \ref{o}, through quantitative analysis, it can be seen that with fixed charge $Q$, as the parameter $\alpha$ increases, both the pericenter and apocenter radii of the particle orbit can maintain stable operation in regions closer to the BH. Similarly, with fixed parameter $\alpha$, increasing the charge $Q$ exhibits the same characteristics. The black dashed lines represent the Schwarzschild BH case.

As can be seen from the right panels of Figs. \ref{n} and \ref{o}, the gravitational waveforms produced by BH with different values of the coupling parameter $\alpha$ and charge $Q$ exhibit notable differences. Specifically, as $\alpha$ or $Q$ increases, the amplitude of the gravitational waveform grows larger, and the phase structure undergoes significant changes. These results indicate that, even within the observationally allowed parameter ranges discussed in Section \ref{3.0}, the corrections due to $\alpha$ and $Q$ may still leave potentially distinguishable characteristic imprints on the gravitational waveforms of periodic orbits in the strong-field regime, thereby providing possible observational signatures for distinguishing among Schwarzschild BH, RNBH, and charged BH in 4D-EGB gravity theory. It should be noted, however, that the present work only addresses the waveform features of individual periodic orbits; for space-borne gravitational wave detectors such as LISA, a more rigorous detectability assessment is generally better suited to be carried out within the framework of long-term EMRI orbital evolution and complete waveform modeling, which we plan to pursue in future work.

\section{Summary and discussion}\label{6.0}

This paper systematically investigates the orbital dynamics and gravitational wave radiation characteristics of charged BH in 4D-EGB gravity theory. As an important representative of quadratic curvature extended gravity, although the construction process of 4D-EGB theory has sparked debates regarding mathematical rigor, subsequent well-defined theories obtained through various regularization approaches have all yielded consistent spherically symmetric BH solutions, providing theoretical support for its physical applicability. Considering the four-dimensional nature of the observed universe, studying the behavior of particles around charged BH and gravitational wave characteristics within the 4D-EGB framework not only helps to reveal the effects of Gauss-Bonnet corrections on BH spacetime structure, but also provides theoretical references for understanding the observational effects of Gauss-Bonnet theory in strong-field regions.

We analyzed the properties of MBO and ISCO. The results show that the orbital radius, angular momentum, and energy of both MBO and ISCO decrease monotonically with increasing coupling parameter $\alpha$ or charge $Q$, and the allowed $(E,L)$ interval of particle bound orbits shifts leftward. Furthermore, using the BH shadow observational data of M87* and Sgr A* as well as the S2 star orbital precession measurement data from the GRAVITY Collaboration, we performed observational constraints on the BH model parameters. The results show that BH shadow observations provide good constraints on both $\alpha$ and $Q$, while the S2 star data only forms effective constraints on the charge $Q \in (0, 0.734)$.

Under fixed energy $E=0.965$ and angular momentum $L=0.5(L_{\text{MBO}}+L_{\text{ISCO}})$, we systematically studied the periodic orbits corresponding to different rational numbers $q$. The results indicate that variations in parameters $\alpha$ and $Q$ significantly affect the energy-angular momentum characteristics of the orbits, and charged BH in 4D-EGB theory exhibit lower energy and angular momentum requirements compared to Schwarzschild BH. The structure of periodic orbits becomes increasingly complex with increasing zoom number $z$ and winding number $w$, and the differences in orbital morphology corresponding to different parameter sets may provide observational avenues for distinguishing gravitational theories. 
 Finally, we computed the gravitational wave waveforms for the (3,1,2) zoom-whirl orbit. The results indicate that, within the observationally allowed parameter ranges considered in this work, larger values of $\alpha$ or $Q$ enable the particle to sustain stable motion in regions closer to the event horizon, leading to enhanced gravitational wave amplitudes and pronounced changes in the phase structure. These theoretically distinguishable waveform features further suggest that, even within the bounds of current observational constraints, the corrections introduced by $\alpha$ and $Q$ may leave discernible imprints on gravitational wave signals in the strong-field regime, thereby providing potential observational signatures for testing 4D-EGB gravity and its deviations from GR through EMRI systems.


It should be noted that this work is built upon the test-particle approximation and a static, spherically symmetric spacetime background, with primary focus on particle orbital dynamics. The gravitational wave radiation analysis is restricted to waveform characteristics within individual orbital periods, and the long-timescale evolutionary process has not yet been addressed. Since dimensionless variables normalized by the BH mass $M$ are adopted throughout, the present results primarily reflect the influence of $Q/M$ and $\alpha/M^2$ on the orbital structure and single-period waveform features. For physical quantities relevant to actual observations, including absolute frequencies, timescales, and long-term evolutionary waveforms, the BH mass $M$, charge parameter $Q$, and Gauss–Bonnet coupling parameter $\alpha$ generally require joint estimation, and the correlations among these parameters as well as potential parameter degeneracies remain to be systematically investigated. Future work may build upon more complete EMRI waveform models and, in conjunction with the sensitivity curves of space-borne gravitational wave detectors such as LISA and TianQin, employ Fisher information matrix analyses and Bayesian parameter estimation methods to further examine the resolvability of these parameters and their mutual correlations. As observational capabilities continue to advance, the present work may provide some reference for observational tests of Gauss–Bonnet gravity.

\section*{Acknowledgements}
We acknowledge the anonymous referee for a constructive report that has significantly improved this paper. 
This work was supported by Guizhou Provincial Basic Research Program (Natural Science) (Grant No.QianKeHeJiChu[2024]Young166), the National Natural Science Foundation of China (Grant No.12365008), the Guizhou Provincial Basic Research Program (Natural Science) (Grant No.QianKeHeJiChu-ZK[2024] YiBan027 and QianKeHeJiChuMS[2025]680), the Guizhou Provincial Major Scientific and Technological Program XKBF (2025) 010 (Hosted by Professor Xu Ning), the Guizhou Provincial Major Science and Technological Program XKGF (2025) 009 (Hosted by Professor Xiang Guoyong) and Guizhou Provincial Major Scientific and technological Program (Teacher Fan Lu Lu moderated).

\newpage
\bibliography{ref}
\bibliographystyle{apsrev4-1}

\end{document}